\input{aipcheck.tex}

  \def\selectedoptions{final}

\documentclass[
   \selectedoptions
  ]
  {aipproc}

\newcommand{\fw} {0.47\textwidth}

\newcommand{\fh} {0.30\textwidth}
\newcommand{\fha}{0.28\textwidth}
\newcommand{\fhb}{0.29\textwidth}
\newcommand{\fhc}{0.28\textwidth}
\newcommand{\fhd}{0.195\textwidth}
\newcommand{\gray}{$\gamma$-ray}
\newcommand{\grays}{$\gamma$-rays}

\newcommand{\pubjournal}[6] {#1, \emph{#2}, {\bf #3}, #4 (#5)}
\newcommand{\pubjournala}[6]{#1, \emph{#2}, #4 (#5)}

\newcommand{\pubprocc}[6]{#1, ``#6,'' in \emph{#2}, #3, #5, #4}
\newcommand{\aap}{A\&A}
\newcommand{\adv}{Adv.\,Spa.\,Res.}
\newcommand{\apj}{ApJ}
\newcommand{\app}{Astropart.\,Phys.}

\newcommand{\icrc}{Int.\,Cosmic\,Ray\,Conf.}
\newcommand{\jcap}{J.\,Cosmol.\,Astropart.\,Phys.}
\newcommand{\jgr}{J.\,Geoph.\,Res.}
\newcommand{\mnras}{Mon.\,Not.\,Roy.\,Astron.\,Soc.}
\newcommand{\physrep}{Phys.\,Rep.}
\newcommand{\pr}{Phys.\,Rev.\,}
\newcommand{\prl}{Phys.\,Rev.\,Lett.}
\newcommand{\rpp}{Rep.\,Progr.\,Phys.}
\newcommand{\ssr}{Spa.\,Sci.\,Rev.}

\layoutstyle{8x11double}

\newcommand\doingARLO[2][]{%
  \ifx\mmref\undefined #1\else #2\fi
}

\begin{document}

\title 
      [Propagation of cosmic rays]
      {Propagation of cosmic rays: nuclear physics\\ in cosmic-ray studies}


\author{Igor V. Moskalenko}{
  address={NASA/Goddard Space Flight Center, Code 661, Greenbelt, MD 20771},
  altaddress={Joint Center for Astrophysics/University of Maryland, Baltimore
County, Baltimore, MD 21250},
  email={imos@milkyway.gsfc.nasa.gov}, thanks={}
}

\author{Andrew W. Strong}{
  address={Max-Planck-Institut f\"ur extraterrestrische Physik,
Postfach 1603, D-85740 Garching, Germany},
  email={aws@mpe.mpg.de}, thanks={}
}

\author{Stepan G. Mashnik}{
  address={Los Alamos National Laboratory, Los Alamos, NM 87544},
  email={mashnik@t2y.lanl.gov},
  homepage={http://www.dcarlisle.demon.co.uk}
}

\copyrightyear  {2001}

\begin{abstract}
The nuclei fraction in cosmic rays (CR) far exceeds the fraction
of other CR species, such as antiprotons, electrons, and positrons. 
Thus the majority of information obtained from CR studies is based
on interpretation of isotopic abundances using CR propagation models
where the nuclear data and isotopic production cross sections in
$p$- and $\alpha$-induced reactions are the key elements.
This paper presents an introduction to the astrophysics of CR
and diffuse \grays\ and discusses some of the puzzles that have emerged 
recently due to more precise data and improved propagation models.
Merging with cosmology and particle physics, astrophysics of CR has
become a very dynamic field with a large potential of
breakthrough and discoveries in the near future.
Exploiting the data collected by the CR experiments to the 
fullest requires accurate nuclear cross sections.

\end{abstract}
\date{\today}
\maketitle

\section{Introduction}

The origin of CR have been intriguing scientists since 1912 when
V.\ Hess carried out his famous balloon flight to measure the ionization
rate in the upper atmosphere.
The energy density of relativistic particles (CR) is $\sim$1 eV cm$^{-3}$
and is comparable to that of the interstellar radiation
and magnetic fields, and turbulent motions of the interstellar
gas. This makes CR one of the essential factors determining
the dynamics and processes in the interstellar medium (ISM).  The observations 
of the Small Magellanic Cloud \citep{sreekumar93} by
the EGRET (Energetic Gamma Ray Experiment Telescope) on board of 
the Compton Gamma Ray Observatory (CGRO) have
shown that the CR are a Galactic and \emph{not} a
``metagalactic'' phenomenon. Observations of the Large Magellanic
Cloud \citep{sreekumar92}, in turn, have shown that \gray\
flux is consistent with CR having a density comparable
to that in our Galaxy.

Major cosmic accelerators are supernova remnants
(SNRs), with a fraction of CR coming from 
pulsars, compact objects in close binary systems, and stellar winds.
Observations of X-ray and \gray\ emission from these objects reveal
the presence of energetic electrons thus testifying to efficient
acceleration \cite{koyama-allen}. 
The total power of Galactic CR sources needed to sustain the observed CR
density is estimated at $5\times10^{40}$ erg s$^{-1}$ which implies
the release of energy in the form of CR of
$\sim$$5\times10^{49}$ erg per supernova (SN) if the SN 
rate is $\sim3$ per century. 

\begin{figure}[!b]
  \resizebox{\fw}{\fhb}{\includegraphics{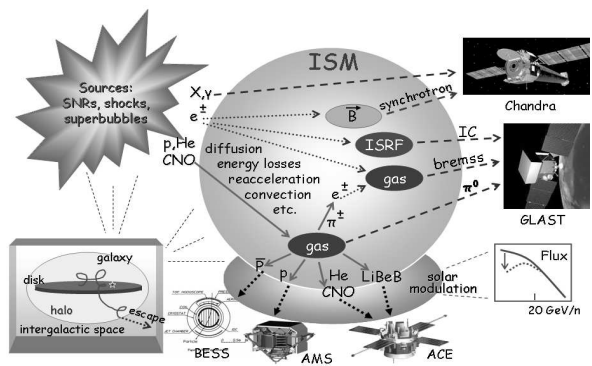}}
\caption{Basic processes in the ISM.}
\label{processes}
\end{figure}

Propagation in the ISM changes the initial composition
and spectra of CR species (Fig.~\ref{processes}). The destruction of
primary nuclei via spallation gives rise to secondary nuclei and
isotopes which are rare in nature, antiprotons, and pions
($\pi^\pm$, $\pi^0$) that decay producing secondary $e^\pm$'s and \grays.
CR are ``stored'' in the Galaxy for tens of millions 
years before escaping into the intergalactic space.

Although much progress has been made since the
direct measurements in space have become possible,
the detailed information refers only to the environment near
to the sun. The CR source composition
and CR propagation history are imprinted in their 
isotopic abundances while diffuse \grays\
and synchrotron emission from different directions
carry clues to the proton and
electron spectra in distant locations. 
These are the only 
pieces of the universal puzzle that we 
have and exploiting them requires extensive modeling.

\section{Indications of new phenomena?}

\begin{figure}[t]
  \resizebox{\fw}{!}{\includegraphics{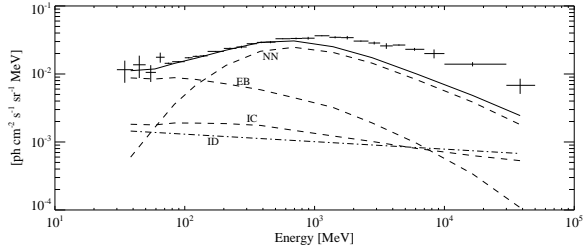}}
\caption{Spectrum ($E^2\times Flux$) of diffuse \grays\ from the inner Galaxy 
as measured by the EGRET.
Curves indicate individual components: $\pi^0$-decay (NN),
electron bremsstrahlung (EB), inverse Compton (IC), and
isotropic diffuse emission (ID). Adapted from \cite{hunter97}.}
\label{hunter}
\end{figure}

The puzzling excess in the EGRET data above 1 GeV (Fig.~\ref{hunter}) 
relative to that
expected \cite{hunter97,M04rev} has shown up in all models that are tuned to
be consistent with local nucleon and electron spectra
\cite{SMR00,SMR04a}. The excess has shown up in all directions,
not only in the Galactic plane. 
An apparent discrepancy between the radial
gradient in the diffuse Galactic \gray\ emissivity and the
distribution of CR sources (SNRs) has worsened the problem 
\cite{SMR00}.

\begin{figure}[b]
  \resizebox{\fw}{\fhd}{\includegraphics{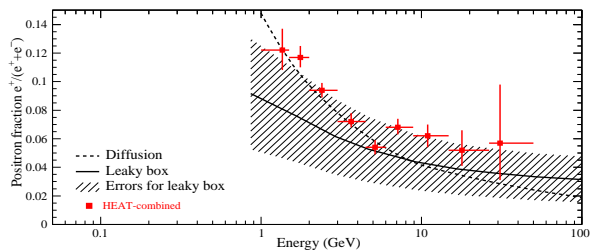}}
\caption{Positrons/(all leptons) ratio in CR compared to calculations
in a leaky-box model \cite{protheroe} (solid) and GALPROP diffusion model 
\cite{MS98} (dashes). Adapted from \cite{coutu}.}
\label{positrons}
\end{figure}

\begin{figure}[t]
  \resizebox{\fw}{\fha}{\includegraphics{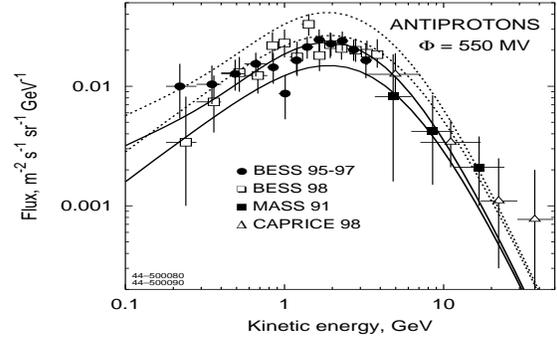}}
\caption{Spectrum of secondary antiprotons in CR as calculated
in reacceleration (solid lines) and optimized (dots) models.
The upper curves -- interstellar, 
lower curves -- modulated ($\Phi=550$ MV) to compare with data.
Adapted from \cite{SMR04a}.}
\label{pbars}
\end{figure}

Positron fraction $e^+$/(all leptons) in CR as measured by HEAT
\cite{heat_positrons} also exhibits an excess above $\sim$8 GeV (Fig.~\ref{positrons}) 
compared to predictions of the diffusion model for secondary production \cite{MS98}. 

Secondary antiprotons are produced in the same interactions 
of CR particles with interstellar gas as $e^+$'s and diffuse \grays.
Recent $\bar p$ data with larger statistics
\cite{Orito00-Maeno01-beach-boezio01-bergstrom} 
triggered a series of calculations of the secondary $\bar p$ 
flux in CR.
The diffusive reacceleration models have certain 
advantages compared to other propagation models: they naturally reproduce
secondary/primary nuclei ratios in CR, have only three
free parameters, and agree better with
K-capture parent/daughter nuclei ratio.
The detailed analysis shows, however, that the reacceleration models 
underproduce $\bar p$'s by a factor of $\sim$2 at 2 GeV \cite{M02} 
(Fig.~\ref{pbars})
because matching the B/C ratio at all energies requires the diffusion
coefficient to be too large. 

If these excesses are not a simple artefact, 
they may be telling us about processes in the ISM, in the Local Bubble,
or signaling exotic physics (e.g., WIMP annihilation, primordial
black hole evaporation), but also may indicate a flaw in the current models.  

\section{Cosmic Rays and Diffuse \grays}
The modeling of CR diffusion in the Galaxy includes the
solution of the transport equation with a given source distribution
and boundary conditions (free escape into the intergalactic
space) for all CR species.
The transport equation describes diffusion and energy losses 
and may also include \cite{Zirakashvili96-seo-ptuskin03}
the convection by a hypothetical Galactic wind,
distributed acceleration in the ISM due to the Fermi 
second-order mechanism (reacceleration), 
and non-linear wave-particle interactions.

The study of transport of the CR nuclear component
requires the consideration of nuclear spallation,
radioactive decay, and ionization energy
losses. Calculation of isotopic abundances involves
hundreds of secondary nuclei produced in
CR interactions with interstellar gas.
A thorough data base of isotopic
production and fragmentation cross sections is thus
a critical element of propagation models
that are constrained by the abundance measurements of isotopes, $\bar p$'s,
and $e^+$'s in CR.

As solar activity changes with a period of 11 years
so does the intensity of CR, but in opposite direction:
with an inverse correlation. The ``solar modulation''
is a combination of effects of convection by the solar
wind, diffusion, adiabatic cooling, drifts,
diffusive acceleration and affects CR below 20 GeV/nucleon.
The theory of solar modulation is far from being
complete \cite{potgieter}. 
The Ulysses spacecraft first provided measurements 
of the solar wind and magnetic field outside the ecliptic
helping us to understanding the global 
aspects of modulation, while
Pioneer and the two Voyagers have explored the outer
solar system. Recently there appear some indications that
Voyager 1, currently at $13.3\times10^9$ km (88 AU) from the sun,
may be close to the termination shock, the heliospheric boundary.
If true, in a few years it will become the first spacecraft ever
to reach interstellar space.

\begin{figure}[t]
  \resizebox{\fw}{!}{\includegraphics{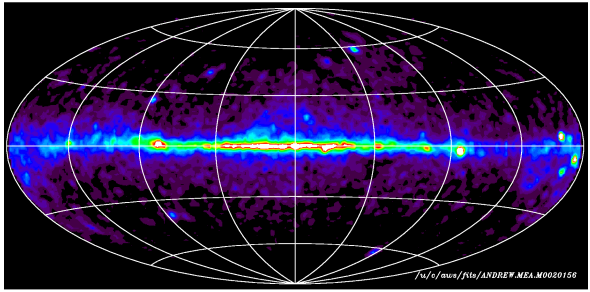}}
\caption{The EGRET sky: diffuse \gray\ emission \cite{M04rev}.}
\label{skymap}
\end{figure}

The diffuse
continuum emission is the dominant feature of the \gray\ sky
(Fig.~\ref{skymap}). 
It is evidence of CR proton and electron
interactions with gas and the interstellar radiation field (ISRF), 
and is created via $\pi^0$-decay, inverse Compton, and bremsstrahlung.
This emission in the range 50 keV
-- 50 GeV has been systematically studied in by OSSE
(Oriented Scintillation Spectrometer Experiment), COMPTEL
(Imaging Compton Telescope), and EGRET on the CGRO and in
earlier experiments \cite{hunter97,M04rev}.
The GLAST (Gamma-ray Large Area Space Telescope) will improve
the sensitivity for the diffuse emission by a factor of 30.

Increasingly accurate balloon-borne and
spacecraft experiments are demanding 
propagation models with improved predictive capability. 
Incorporation of the realistic astrophysical input increases the
chances of the model to approach reality and dictates that 
such a model should be numerical. Besides the nuclear data,
such a model has to include the detailed 
3-dimensional maps of the Galactic gas derived from radio and IR surveys, the
Local Bubble and local SNRs, the spectrum of the ISRF,
magnetic fields, details of composition of interstellar dust, grains, 
as well as theoretical works on CR acceleration and 
transport in Galactic environments. 
The most advanced model to date, GALPROP, is a three dimensional model
\cite{SMR00,M02,SM98}. It is widely used as a 
basis for many studies, such as search for dark matter
signatures, origin and evolution of elements, the spectrum and origin of Galactic
and extragalactic diffuse \gray\ emission, heliospheric modulation.
The model calculates CR propagation for nuclei 
($_1^1$H to $_{28}^{64}$Ni), $\bar p$'s, $e^\pm$'s, and 
computes \grays\ and synchrotron emission in the same framework;
it includes all relevant processes and reactions.
It is an excellent tool to cross-test various hypotheses \cite{SMR04a,MSR98}.

\begin{figure}[t]
  \resizebox{\fw}{\fh}{\includegraphics{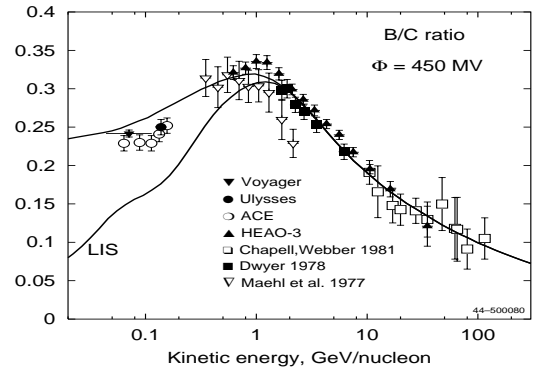}}
\caption{Boron/carbon ratio in CR as calculated in the reacceleration
model. Lower curve -- interstellar ratio, upper curve -- modulated ($\Phi=450$
MV) to compare with data. Adapted from \cite{SMR04a}.}
\label{bc}
\end{figure}

\section{Nuclear physics in CR studies}

The abundances of stable ($_3$Li, $_4$Be, $_5$B, 
$_{21}$Sc, $_{22}$Ti, $_{23}$V) and radioactive secondaries 
($^{10}_{4}$Be, $^{26}_{13}$Al, $^{36}_{17}$Cl, $^{54}_{25}$Mn) in CR
are used to derive the diffusion 
coefficient and the halo size \cite{SM98,ptuskin-webber}.
The derived source abundances of CR
may provide some clues to mechanisms and sites of CR acceleration.
However, the interpretation of CR data, e.g.,
the sharp peak in the secondary/primary nuclei 
ratio (Fig.~\ref{bc}), is model dependent.
The leaky-box model fits the secondary/primary ratio by 
allowing the path-length distribution vs.\ rigidity to vary.
The diffusion models are more physical and explain the shape of
the secondary/primary ratio in terms of
diffusive reacceleration (distributed energy gain) in the ISM, 
convection by the Galactic wind,
or by the damping of the interstellar turbulence by 
CR on a small scale.

The secondary/primary nuclei ratio is sensitive to the 
value of the diffusion coefficient and its energy dependence.
A larger diffusion coefficient leads to a
lower ratio since the primary nuclei escape faster from the Galaxy 
producing less secondaries and vice versa.
The abundance of radioactive secondaries
(e.g., $^{10}$Be/$^9$Be) is sensitive to the Galactic halo size, the 
Galactic volume filled with CR. The larger the halo the longer
it takes for radioactives to reach us thus decreasing the ratio
$^{10}$Be/$^9$Be.

\begin{figure}[t]
  \resizebox{\fw}{\fha}{\includegraphics{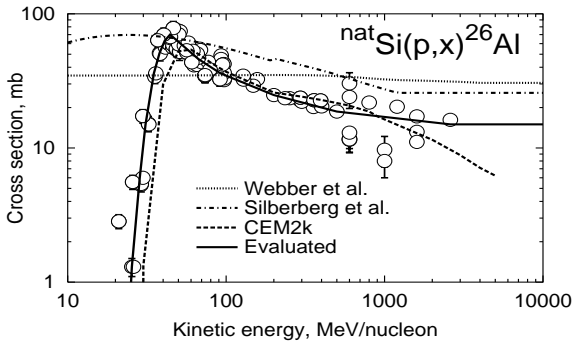}}
\caption{Cross section for the reaction $^{nat}$Si$(p,x)^{26}$Al.
Calculations: semi-empirical systematics \cite{webber_cs,ST}, 
CEM2K \cite{mashnik04}, evaluated \cite{MMS01}. 
Adapted from \cite{mashnik04}.}
\label{Al}
\end{figure}

\begin{figure}[b]
  \resizebox{\fw}{!}{\includegraphics{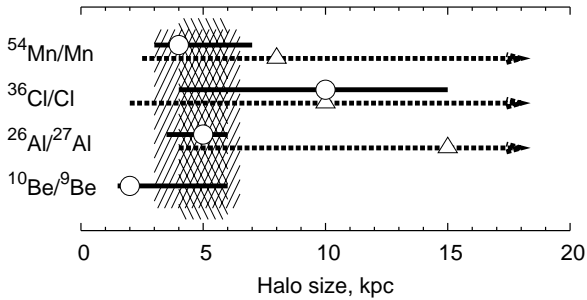}}
\caption{Determination of the Galactic halo size based on radioactive 
secondaries: using semi-empirical systematics \cite{SM01} -- dashes,
light shaded area shows the range consistent with all ratios;
using evaluated cross sections -- solid, heavy shaded area shows the
range consistent with all ratios. Adapted from \cite{MMS01}.}
\label{halo}
\end{figure}

Current CR experiments,
such as Advanced Composition Explorer (ACE), Ulysses, and
Voyager, deliver excellent quality spectral and isotopic data.
Meanwhile, isotopic production cross sections have for a long time 
been the Achilles' heel of CR propagation models, and now
become a factor restricting further progress.
Interpretation of CR data requires massive calculations
of isotopic production involving $p$'s and $\alpha$'s, however,
the widely used semi-empirical systematics are 
frequently wrong by a significant factor \cite{MMS01,yanasak-MM03};
this is reflected in the value of propagation parameters.
Figs.~\ref{Al} and \ref{halo} illustrate the effect of isotopic
cross sections on the derivation of the halo size from radioactive
secondaries. Using the semi-empirical systematics leads to
the huge error bars where the upper limits are
consistent with infinite halo size \cite{SM01}. 
Using the evaluated cross sections
dramatically reduces the error bars and gives a 
consistent value: 4--6 kpc \cite{MMS01}.
Unfortunately, the evaluated cross sections in the required energy
range are mostly unavailable.

\begin{figure}[t]
  \resizebox{\fw}{\fha}{\includegraphics{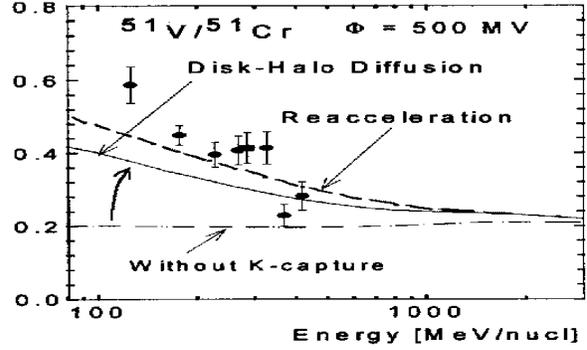}}
\caption{$^{51}$V/$^{51}$Cr ratio in CR as calculated with/without
K-capture. Adapted from \cite{jones01b}.}
\label{K-cap1}
\end{figure}

\begin{figure}[b]
  \resizebox{\fw}{\fhc}{\includegraphics{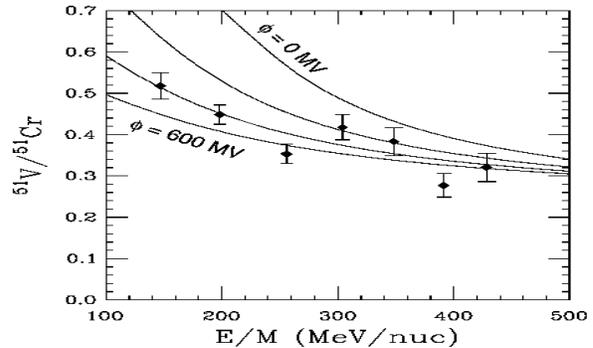}}
\caption{$^{51}$V/$^{51}$Cr ratio in CR as calculated for
different levels of solar modulation. Interstellar 
ratio corresponds to $\Phi=0$ MV. Adapted from \cite{niebur03}.}
\label{K-cap2}
\end{figure}

K-capture isotopes in CR 
(e.g., $^{49}_{23}$V, $^{51}_{24}$Cr) can 
serve as important energy markers and can be used to study
the energy-dependent effects such as diffusive 
reacceleration in the ISM and heliospheric
modulation \cite{soutoul98,jones01b,niebur03}.
Such nuclei usually decay via electron-capture and have a short lifetime
in the medium. In CR they are stable or live longer
as they are created bare by fragmentation
of heavier nuclei while their $\beta^+$-decay mode is suppressed.
At low energies, their lifetime depends on the balance between
the processes of the electron attachment from the ISM
and stripping. The probability of attachment is 
strongly energy-dependent, increasing toward low energies, while
the probability of stripping is flat. This 
makes the abundances of K-capture isotopes in CR energy-dependent
(Fig.~\ref{K-cap1}). Without K-capture, 
the ratio $^{51}$V/$^{51}$Cr in CR would be flat because both
nuclei are secondary. The electron K-capture
$^{51}$Cr(EC)$^{51}$V increases the ratio at low energies. 
Reacceleration (energy gain)
increases it even further. On the contrary, solar modulation 
flattens the ratio (Fig.~\ref{K-cap2}), which is very steep in the ISM.

Study of the light nuclei in CR (Li--O) allows
us to determine propagation parameters
averaged over a larger Galactic region, but the local ISM 
is \emph{not} necessarily the same and the \emph{local} propagation 
parameters may significantly differ. The best way to study the local
ISM is to look at isotopes with shorter lifetimes 
(e.g., $^{14}$C) and heavy nuclei since
large fragmentation cross sections lead to a small ``collection area.''

\begin{figure}[t]
  \resizebox{\fw}{\fh}{\includegraphics{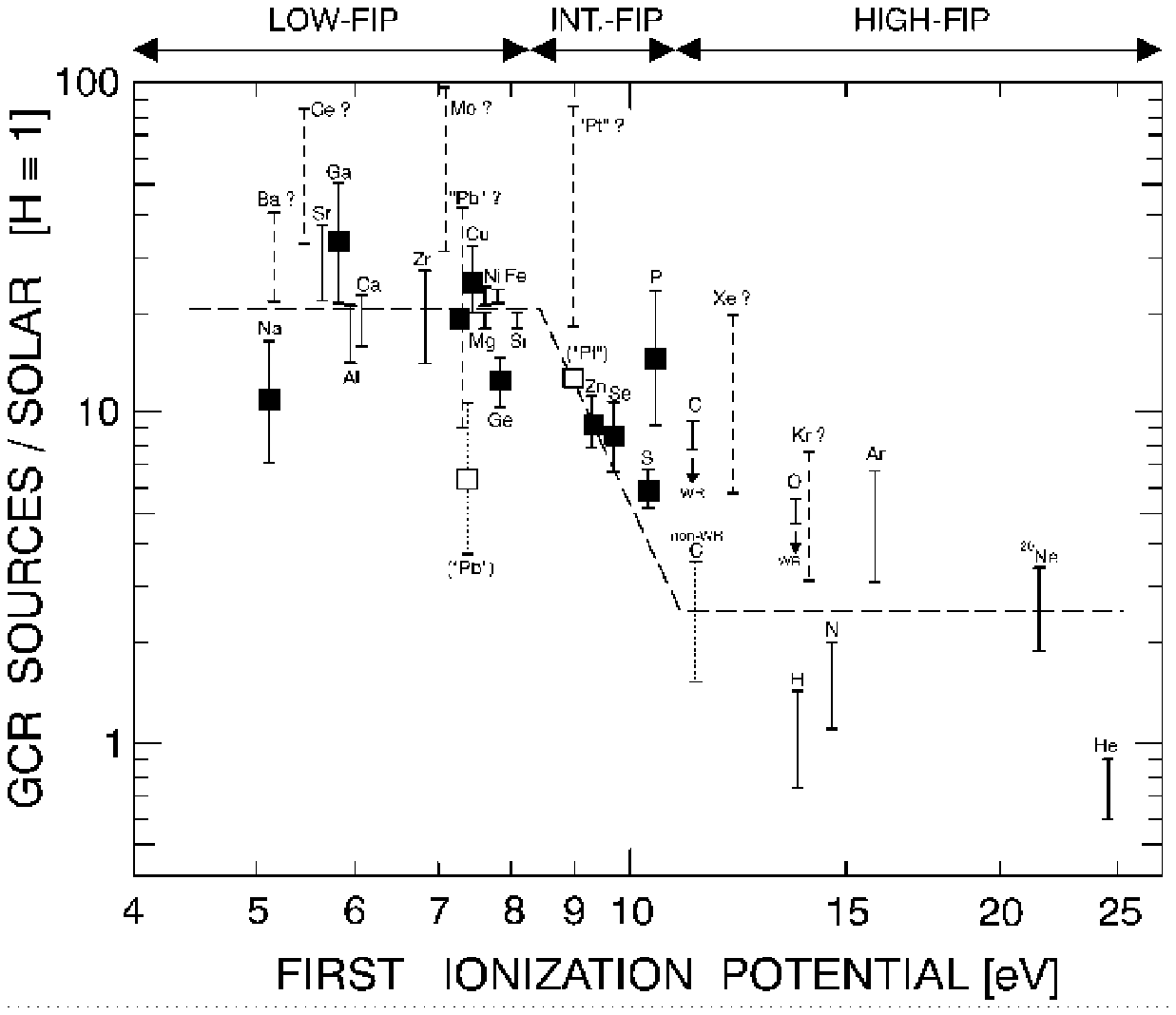}}
\caption{(Galactic CR sources)/(solar system) abundances of different 
elements vs.\ FIP. Adapted from \cite{meyer97}.}
\label{fip}
\end{figure}

The CR source composition is derived from direct CR data
by correcting for the effects of propagation, spallation, and solar
modulation.
The elements with low first-ionization potential (FIP) appear to be
more abundant in CR sources relative to the high-FIP 
elements, when compared with the solar system material (Fig.~\ref{fip}). 
This might imply that the source material for
CR includes the atmospheres of stars with 
temperatures $\sim$$10^4$ K \citep{casse78}.
A strong correlation between FIP and ``volatility''
(most of low-FIP elements are
refractory while high-FIP elements are volatile) suggests that
CR may also originate in the interstellar dust,
pre-accelerated by shock waves \citep{meyer97,epstein80}. $_{11}$Na,
$_{31}$Ga, $_{37}$Rb and some other elements $Z>28$ break 
this correlation. CR data tend to prefer volatility over FIP, but 
uncertainties in the derived source abundances (cross
sections!) prevent an unambiguous solution.

Isotopic peculiarities of CR composition are also important.
For example, the $^{22}_{11}$Ne/$^{20}_{11}$Ne 
enrichment might tell us that CR are produced
in cores of superbubbles \cite{higdon03} created by multiple 
correlated SNe.

\section{Scent of new physics}

Galactic and extragalactic space presents a test range 
where nature runs its numerous experiments continuously for billions
of years. This is an arena where all fundamental forces perform
in an exotic show involving yet-to-be-discovered particles, 
new elements, giant nuclei bound by gravitation
-- neutron stars, and singularities -- black holes, 
and engineering the largest-scale grid of structures in the universe.
CR and diffuse \grays, therefore, could contain signatures 
of exotic physics,
however, conventional CR present an enormous background for tiny 
exotic signals.

The growing number of experiments forces us to the conclusion that
the universe is dominated by the dark matter (DM) and dark energy. A preferred candidate 
for non-baryonic DM is a weakly interacting
massive particle (WIMP). The WIMP is the lightest neutralino $\chi^0$
\cite{jkg-bergstrom00}, which arises in supersymmetric
models of particle physics, or 
a Kaluza-Klein hypercharge $B^1$ gauge boson \cite{matchev}. 
Annihilation of neutralinos creates a soup of
particles, which eventually decay to ordinary baryons and leptons.
The DM particles in the halo
or at the Galactic center \cite{gunn-stecker-silk-gondolo-silk-gondolo} 
may thus be detectable via their annihilation products 
($e^+$, $\bar p$, $\bar d$, \grays) in CR 
\cite{bergstrom99-baltz99-baltz03-darksusy04}. The approach is to scan the
SUSY parameter space to find a candidate able to fill the 
excesses in diffuse \grays, $\bar p$'s, and
$e^+$'s over the predictions of a conventional model (as discussed above).
Preliminary results of the ``global fit'' to the 
$e^+$'s, $\bar p$'s, and diffuse \gray\ data simultaneously 
look promising \cite{deboer1-deboer2}. In particular, the
DM distribution with two-ring structure
allows the EGRET \gray\ data and the rotation curve of the Milky Way
to be fitted, while the ring radii, 5 and 14 kpc, surprisingly well 
coincide with the observed rings of cold H$_2$ gas and stars, respectively.

\begin{figure}[t]
  \resizebox{\fw}{\fha}{\includegraphics{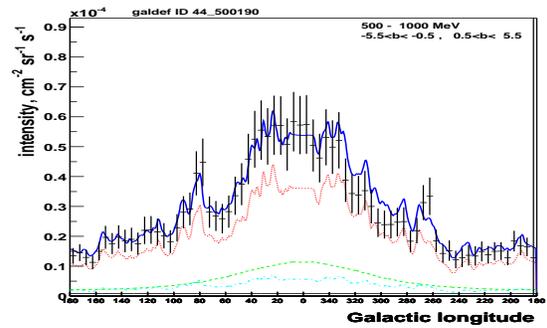}}
\caption{Longitude profile of diffuse \gray\ emission \cite{SMR04a}.
Lines (top to bottom): total, $\pi^0$-decay, inverse Compton, bremsstrahlung,
extragalactic background.}
\label{longitude}
\end{figure}

\begin{figure}[t]
  \resizebox{\fw}{\fha}{\includegraphics{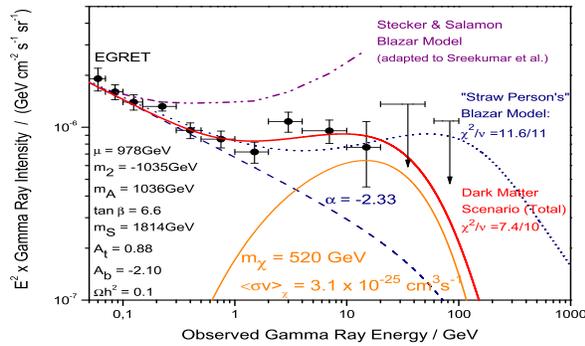}}
\caption{Spectrum of extragalactic diffuse \gray\ emission \cite{SMR04b}. 
Adopted from \cite{mannheim}.}
\label{egrb}
\end{figure}

In terms of conventional
physics, the spatial fluctuation of CR intensity may also
provide a feasible explanation. The CR $\bar p$ data 
can be used to derive the Galactic
average proton spectrum (Fig.~\ref{pbars}, optimized model), 
while the electron spectrum is adjusted using
the diffuse \grays\ \cite{SMR04a}.
The model shows a good agreement with EGRET spectra of diffuse \gray\
emission ($<$100 GeV) from different sky regions (Fig.~\ref{longitude}).
The increased Galactic contribution to the diffuse emission reduces
the estimate of the extragalactic \gray\ background \cite{SMR04a,SMR04b}.
The new extragalactic background \cite{SMR04b} shows a positive
curvature (Fig.~\ref{egrb}), 
which is expected if the sources are unresolved blazars
or annihilations of the neutralino DM \cite{salamon-ullio,mannheim}. 
The discrepancy between the radial
gradient in the diffuse Galactic \gray\ emissivity and the
distribution of SNRs can be solved \cite{SMR04c}
if the ratio H$_2$/CO in the ISM increases
from the inner to the outer Galaxy. 
The latter is expected from the Galactic metallicity gradient. 

The difficulty associated with $\bar p$ may also indicate new effects. 
The propagation
of low-energy particles may be aligned to the magnetic field lines 
instead of isotropic diffusion \cite{M02}. Our local environment (the Local 
Bubble) may produce a fresh ``unprocessed'' nuclei component in CR
at low energy \cite{M03}; the evidence for SN
activity in the solar vicinity in the last few Myr supports this idea.

\section{Conclusion}

Several topics are expected to become the subject of intensive 
studies in the coming years.
PAMELA (Payload for Antimatter Matter Exploration and Light-nuclei
Astrophysics) is designed to measure $\bar p$'s,
$e^\pm$'s, and isotopes H--C over 0.1--300 GeV. 
Future Antarctic flights of a new BESS-Polar instrument
(Balloon-borne Experiment with a Superconducting Spectrometer)
will considerably increase the accuracy of data on $\bar p$'s
and light elements. The AMS (Alpha Magnetic Spectrometer)
onboard the International Space Station
will measure CR particles and nuclei
$Z\hbox{\rlap{\hbox{\lower3pt\hbox{$\sim$}}}\lower-2pt\hbox{$<$}}26$
from GeV to TeV energies. This is complemented by measurements of
heavier nuclei $Z>29$ by (Super-)TIGER (Trans-Iron Galactic Element Recorder).
The future GLAST mission will be capable of measuring $\gamma$-rays in the range 20
MeV -- 300 GeV; besides other goals, it should deliver a final
proof of proton acceleration in SNRs -- long awaited by the CR
community. A breakthrough on SUSY and high-energy interactions 
should come with operation of the new CERN large hadronic collider, LHC.
Not surprisingly,
the success of the state-of-the-art CR experiments depends heavily 
on the quality of nuclear data and especially $p$- and 
$\alpha$-induced reactions at intermediate energies from 
tens of MeV -- few GeV. Challenging these new frontiers is thus impossible 
without involvement of the nuclear physics community.

This work was supported in part by a NASA Astrophysics Theory Program grant,
US DOE, and the CRDF Project MP2-3025.

\bibliographystyle{aipproc}

\end{document}